\documentstyle[prl,aps]{revtex}
\input epsf
\def\a{a}

\def\t{\tau}
\def\ba{\begin{eqnarray}}
\def\ea{\end{eqnarray}}
\def\be{\begin{equation}}
\def\ee{\end{equation}}

\def\ie{{\em i.e.\/}}  
\def\eg{{\em e.g.\/}}

\def\E{{\mathcal E}}  
\def\S{{\mathcal S}}  
  
\def\L{{\mathcal L}}

\def\N{{\mathcal N}}

\def\half{{1\over2}}  

\def\be{\begin{equation}}
\def\ee{\end{equation}}
\def\bea{\begin{eqnarray}}
\def\eea{\end{eqnarray}}
\def\bs{\begin{subequations}}
\def\es{\end{subequations}}
\def\g{\gamma}

\def\vp{\varphi}

\def\d{\partial}

\def\e{\epsilon}

\def\vp{\varphi}

\def\s{\sigma}
\def\t{\tau}
\def\a{\alpha}
\def\b{\beta}

\newcommand{\eq}[1]{equation~(\ref{#1})}

\newcommand{\fig}[1]{Fig.~(\ref{#1})}

\def\mxth{\mathsurround=0pt }
\def\xversim#1#2{\lower2.pt\vbox{\baselineskip0pt \lineskip-.2pt
    \ialign{$\mxth#1\hfil##\hfil$\crcr#2\crcr\sim\crcr}}}

\def\Montreal{Montr\'eal}  
\def\Quebec{Qu\'ebec}

\begin{document}

\twocolumn[\hsize\textwidth\columnwidth\hsize\csname@twocolumnfalse\endcsname

\title{Hawking radiation of nonsingular black holes in two dimensions
 }

\author{D. A. Easson}

\address{
Department of Physics, 
McGill University,
\Montreal, \Quebec, H3A 2T8, Canada \\
\date{\today}
e-mail: easson@hep.physics.mcgill.ca
}
\maketitle

\begin{abstract}
In this letter we study the process of Hawking radiation of a black hole assuming the existence
of a limiting physical curvature scale. The particular model is constructed using 
the Limiting Curvature 
Hypothesis (LCH) and in the context of two-dimensional dilaton gravity. The black hole 
solution exhibits properties of the standard Schwarzschild solution at large 
values of the radial coordinate.
However, near the center, the black hole is nonsingular and the metric becomes 
that of de Sitter spacetime. The Hawking temperature is calculated using 
the method of complex paths.
We find that such black holes radiate eternally and never completely
evaporate. The final state is an eternally radiating relic, near the fundamental scale,
which should make a viable dark matter candidate. We briefly comment on the
black hole information loss problem and the production of such black holes in collider experiments.
\\
\vspace{.2cm}
PACS number(s): 04.60.-m; 04.70.Dy; 98.80.Cq. \hspace{1.7cm} MCGILL-02-32, \,\, \tt hep-th/0210016\rm
\end{abstract}]
\section{Introduction}  
\label{intro}  
Two of the outstanding questions in physics today are how should Einstein's
theory of gravity be modified in regions of spacetime with high curvatures, and what is
the final state of an evaporating black hole. The physics required to answer
both of these questions rests in an as of yet unknown theory of quantum gravity. 

In general, theories which report to describe features of quantum gravity (including some string theories) 
contain higher derivative corrections to the Einstein-Hilbert action \cite{'tHooft:bx} - \cite{Fradkin:ys}.
For example, perturbative string theory predicts an infinite series of correction terms to the
Einstein equations
\be
R_{\a\b} - \half g_{\a\b} R + {\mathcal O} (\a^\prime R^2) + {\mathcal O} (\a^{\prime 2} R^4) + \cdots =0
\,.
\ee
These correction terms become important at energies near the
string scale $\sqrt{\a^\prime}=M_s^{-1}$.
It is possible that such terms will drastically affect the structure of singularities in both 
cosmological and black hole spacetimes. Perhaps quantum gravity will some how remove
the singularities, leading to geodesically complete spacetimes. 
 
Two examples of higher-derivative gravitational theories which
eliminate the singularity in the Pre-Big-Bang model (allowing for a successful graceful exit) are 
constructed in~\cite{Brandenberger:1998zs,Easson:1999xw}. These models utilize the Limiting Curvature Hypothesis (LCH)
of~\cite{markov}. The goal of the LCH is to capture
features of quantum gravity in an effective theory which interpolates between General Relativity at low
energies and a nonsingular spacetime at high curvatures. The LCH has been studied in a variety of
black hole and cosmological spacetimes \cite{Brandenberger:1998zs} - \cite{Brandenberger:1995es}
(also see~\cite{Frolov:pf} - \cite{Morgan:1990yy}).  The assumption of the existence of a limiting curvature is 
well motivated given the existence of a fundamental length scale $\ell_f$ (such as the Planck length $\ell_{pl}$ 
or the string length $\ell_{s}$).  If there is a fundamental length then it follows
(from dimensional arguments) that all curvature invariants are bounded
\ba
|R|<\ell_{f}^{-2},\quad |R_{\a\b}R^{\a\b} | < \ell_{f}^{-4}, \\
\quad |C_{\a\b\g\delta}C^{\a\b\g\delta}| < \ell_{f}^{-8}
\,,\dots
\ea

In this letter we investigate the process of Hawking radiation of a nonsingular black hole 
in $(1+1)$-dimensional dilaton gravity~\cite{Trodden:1993dm}. The construction involves concepts
from the LCH. However, since the Ricci scalar is the only invariant in two dimensions, the role
of the higher derivative correction terms is facilitated by the dilaton and its potential. The resulting
solution is nonsingular everywhere and resembles the Schwarzschild black hole at large values of
the radial coordinate. At $r=0$ the solution approaches the de Sitter metric. Cosmologically desirable
consequences may arise if the universe is generated at the interior of a black hole~(see, \eg \cite{Easson:2001qf}
and the references therein).

Although it is natural to question the physical validity of results in $2D$ dilaton gravity,
there are strong motivations for using the theory as a toy model of quantum gravity
and to study the process of black hole evaporation. In particular, a certain limit of superstring theory
yields a generalized dilaton theory as the effective action (see \eg~\cite{Grumiller:2002nm}). This simplified 
model allows for completely analytic calculations, possibly capturing non-perturbative features of
the physics, evaporation and final state of realistic four-dimensional black holes.~\footnote{For some
early references to two-dimensional gravity see \cite{D'Hoker:1982er} - \cite{Kazakov:2001pj}.}  Further motivation
is provided by \cite{Youm:1999xn}, where it is argued that the thermodynamics of black hole in $D \geq 4$
is effectively described by the thermodynamics of black holes in two-dimensional dilaton gravity theories.
Specifically, it is shown that the Bekenstein-Hawking (BH) entropies of single-charged dilaton black holes and
dilaton $p$-branes with an arbitrary dilaton coupling parameter in arbitrary 
spacetime dimensions are exactly reproduced by the BH entropy of the two-dimensional black 
hole in the associated $2D$ dilaton gravity model.  We are also inspired by the successful generalization
of a nonsingular cosmological model in two dimensions \cite{Moessner:1994jm} to a nonsingular 
four-dimensional model~\cite{Brandenberger:1998zs,Easson:1999xw} using dilaton gravity and the 
limiting curvature construction mentioned above.  Based on this
encouraging evidence, we believe that the two-dimensional results presented in this paper should
apply to more realistic four-dimensional models.

\section{A nonsingular black hole}
\label{nons}  
Here we reconstruct the nonsingular two-dimensional dilaton
black hole solution presented in~\cite{Trodden:1993dm}.
We begin with the most general Lagrangian for gravity with a scalar field 
(\eg, dilaton gravity) in $(1+1)$-dimensions~\cite{Banks:1992xs}:
\be
\L = \sqrt{-g} \left( D(\varphi) \, R + G(\varphi)\,(\nabla\varphi)^2 + H(\varphi)\right)
\,.
\ee
By performing a conformal transformation
\be
g_{\a\b} \rightarrow e^{2 \s(\vp)} g_{\a\b}
\,,
\ee
and requiring that 
\be
4\frac{d\s}{d\vp} \frac{dD}{d\vp} = - G(\vp)
\,,
\ee
it is possible to conformally transform away the kinetic term for $\vp$ and we may simply
start with the action
\be\label{start}
S=\int d^2x \sqrt{-g} \left(D(\vp) \,R + V(\vp) \right)
\,,
\ee
where $V(\vp)=e^{2 \s(\vp)}H(\vp)$.~\footnote{Note that in~\cite{Grumiller:2000wt}, it is
argued that under certain circumstances conformally related theories may not be equivalent.}

Varying this action with respect to $\vp$ and the metric tensor yields
\bea
- \frac{\d V(\vp)}{\d \vp} &=& \frac{\d D(\vp)}{\d \vp} R \label{eo1} \\
V(\vp)\,g_{\a\b} &=& 2(\nabla^2 g_{\a\b} - \nabla_\a \nabla_\b)D(\vp)
\,,
\eea
respectively.  We may redefine $\vp$ so that $D(\vp)=1/\vp$ and assume a spherically symmetric and
static metric 
\be\label{ds2d}
ds^2 = -n(r)dt^2 + p(r)dr^2
\,.
\ee
In the ``Schwarzschild gauge" $p(r)=1/n(r)$ and the EOM become
\bea\label{eom}
\vp^3 V(\vp) - 4 n \vp'^2 + n'\vp'\vp + 2 n \vp \vp'' &=& 0 \\
\frac{\d V }{\d \vp} + \vp^{-2} n'' &=& 0 \\
V(\vp) + \vp^{-2} \vp' n' &=& 0
\,.
\eea
Implementing the limiting curvature hypothesis is equivalent to identifying a class of potentials 
$V(\vp)$ so that $R$ is bounded for all values of the radial coordinate from
$0$ to $\infty$.  
The form of the potential is restricted by considering the behavior of the system at low and high curvatures.  
We require $n(r)$ to reduce to the Schwarzschild solution in regions of spacetime with small curvature.  
Mathematically,
\be
n(r)\rightarrow (1-2m/r) \qquad when \qquad r\rightarrow \infty, \; \vp \rightarrow 0
\,.
\ee
From \eq{eo1} it is easy to see that
limiting $V'(\vp)$ in the high curvature regime (\ie \, $r \rightarrow 0$, $\vp \rightarrow \infty$)
will result in the limitation of $R$.

While there are many potentials that interpolate correctly between the above asymptotic regimes,
a simple potential which satisfies the above criteria is
\be
V(\vp) = \frac{2 m A^3 \vp^2}{1 + m A^3 l^2 \vp^3}
\,,
\ee
where $A$ is a constant and $l$ is a constant representing the limiting fundamental scale.~\footnote{
Note that this potential is only one possible potential that will force the curvature to be limited.
While it is not a unique choice all potentials that satisfy the asymptotic conditions, limiting the 
curvature, will give the same result.  For a more detailed explanation of
the construction of $V(\vp)$ we refer the reader to the original source~\cite{Trodden:1993dm}.}
Using this potential in the EOM one finds the exact solutions
\bea
\vp(r) = \frac{1}{Ar}\,,\qquad \qquad \qquad \qquad \qquad \;\;\;\;\;\;\\
n(r) =  \frac{1}{3} \Big(\frac{m}{l}\Big)^{2/3}\,\times \qquad \qquad \qquad \qquad \nonumber\\
\ln{\Big\{\frac{r^2 - (ml^2)^{1/3}r + \nonumber
(ml^2)^{2/3}}{r_0^2 - (ml^2)^{1/3}r_0 + (ml^2)^{2/3}} \Big(\frac{r_0 + (ml^2)^{1/3}}{r + (ml^2)^{1/3}}\Big)^2 \Big\}} \nonumber \\
	+ \frac{2}{\sqrt{3}} \Big(\frac{m}{l}\Big)^{2/3}\,
	\Big\{arctan\Big(\frac{2r - (ml^2)^{1/3}}{\sqrt{3} (ml^2)^{1/3}}\Big) \nonumber \\
	-arctan\Big(\frac{2r_0 - (ml^2)^{1/3}}{\sqrt{3} (ml^2)^{1/3}}\Big)\Big\} \label{nofr}
\,,
\eea
where $r_0$ is the location of the black hole horizon.  The function $n(r)$ is plotted below
in~Fig. (I).  When the mass of the black hole is large, 
$n$ reduces to $(1-2m/r)$ as expected.

\begin{figure}\label{f.eps}
\begin{center}
 \epsfxsize=3.2 in \centerline{\epsfbox{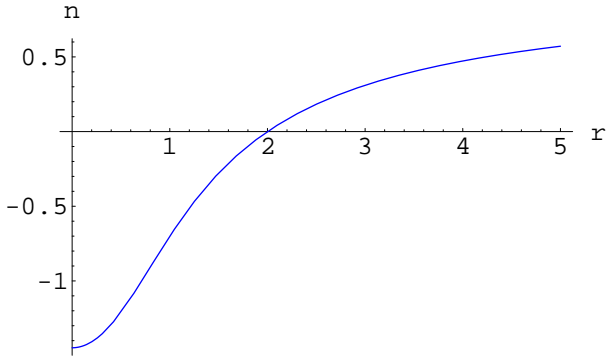}}
  \end{center}
    \caption{
A plot of $n(r)$ for the values $l=m=1$. (These values are
used in the plots throughout the paper.)  
Far from the point $r=0$ the solution
is indistinguishable from the Schwarzschild solution. The function $n$ vanishes at the
horizon $r=r_0=2m$. Unlike the Schwarzschild solution, $n$ remains finite at $r=0$.}
		    \end{figure}

The Ricci scalar $R$ (the only curvature invariant in two-dimensions) remains finite for all
positive value of the radial coordinate $r$,
\be\label{eqR}
R=\frac{2m(2r^3 - l^2 m)}{(r^3 + l^2 m)^2}
\,,
\ee
hence, the spacetime is singularity free (see~\fig{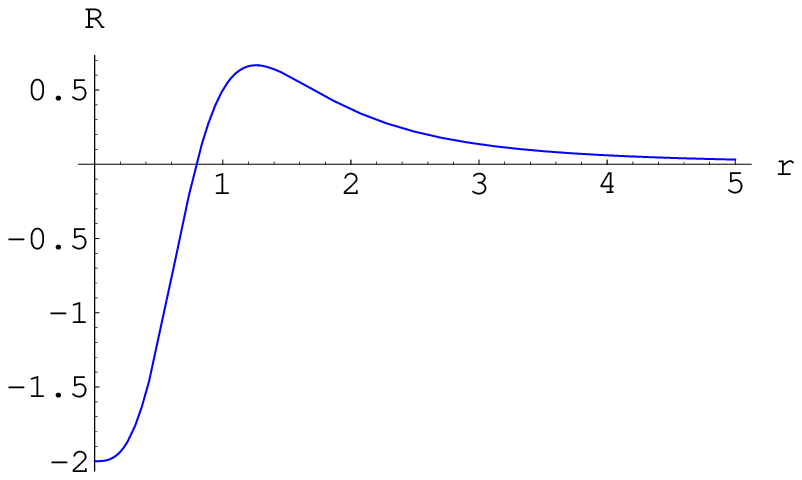}).
Note that $R$ changes sign at the value $r=(l^2m/2)^{2/3}$ and takes on its limiting value
$R_{max}$ at $r=0$,
\be\label{rmax}
R_{max} = -\frac{2}{l^2}
\,.
\ee
\begin{figure}\label{R.eps}
\begin{center}
 \epsfxsize=3.2 in \centerline{\epsfbox{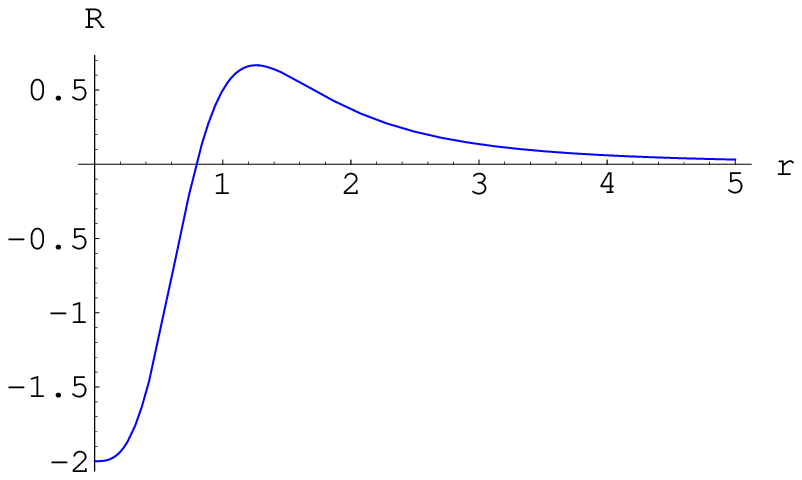}}
  \end{center}
    \caption{The curvature $R$ remains finite for all
	$r$ from $r=0$ to $r=\infty$.}
		    \end{figure}

Near $r=0$ the spacetime becomes de Sitter.~\footnote{In the conventions of~\cite{Trodden:1993dm}, the cosmological 
constant ($\Lambda \propto R$) for de Sitter space is
negative.}
To see this, expand $n$ around the point $r=0$:  
\be
n(r)\simeq -C + \frac{r^2}{l^2} + {\mathcal{O}}[(r)^3]
\,.
\ee

Ignoring terms of order $r^3$ and higher, the line element becomes
\be
ds^2 = -\left(\frac{r^2}{l^2} - C \right)dt^2 + \left(\frac{r^2}{l^2} - C \right)^{-1}dr^2
\,,
\ee
where $C$ is a constant.  Taking $C=1$, we may define a new coordinate $\t$ given by
\be
\frac{d\t}{dr} = \Big((\frac{r}{l})^2 -1 \Big)^{-1/2}
\,,
\ee
which gives
\be
\frac{\t(r)}{l} = arccosh\left(\frac{r}{l}\right)
\,.
\ee
In the $(t,\t)$ coordinates the metric takes on the recognizable de Sitter form
\be
ds^2 = d\t^2 - sinh^2\left(\frac{\t}{l}\right)dt^2
\,.
\ee

Another conceivably interesting region is near the fundamental scale $r\simeq l$.
Expanding around $r=l$, $n$ takes the form
\be
n(r)\simeq C' + \frac{2m}{l(l+1)} (r-l) + {\mathcal{O}}[(r-l)^2]
\,.
\ee
Here we choose a new coordinate $\chi$ (and $C'=1$) defined by
\be
\frac{d\chi}{dr}=(\frac{2m(r-l)}{l(l+1)} - 1)^{-1/2}
\,.
\ee
In the $(t,\chi)$ coordinates the metric becomes
\be
ds^2 = d\chi^2 - \a^2 \chi^2 dt^2
\,,
\ee
where the constant $\a = m/l(l+1)$.  This is simply the Milne metric.

One final comment is in order concerning the above solution. In~\cite{Horowitz:1995ta},
Myers and Horowitz argue that spacetime singularities should play a useful role
in gravitational theories by eliminating unphysical solutions. Furthermore, they
suggest that any modified theory of gravity must have singularities in order to
posses a stable ground state. The argument goes as follows: even if the theory claims to have a ground
state with $E<0$, one can always start with the Schwarzschild metric with $m<E$ and argue that it must 
be singular. Removing all singularities leads to the existence of states with arbitrarily negative energy.
They suggest that a new mechanism must be found to prevent a state
which resembles the negative mass Schwarzschild solution from existing in the theory.

In the above model, we may argue that the negative mass Schwarzschild solution is not a well behaved solution,
in the sense that a singularity is present for $m<0$ (see \eq{eqR}).  
Furthermore, we may argue that the negative energy solutions may simply be thrown out due to their 
pathological, singular nature.
Negative mass Schwarzschild solutions are not possible asymptotic
solutions of the theory since a singularity is
present at finite $r$. Thus, the instability of the
theory for $m < 0$ discussed in~\cite{Horowitz:1995ta} does not
occur in our theory.~\footnote{I would like to thank R.~Myers for useful discussions concerning this point.}

\section{Hawking radiation and the method of complex paths}
\label{hawk}  
We now examine the process of Hawking radiation of the nonsingular
black hole solution obtained in the previous section.  One of the simplest
ways to do this is to employ the method of complex path analysis introduced by
Landau~\cite{Landau:qm} and applied to Schwarzschild-like spacetimes by Srinivasan and 
Padmanabhan~\cite{Srinivasan:1998ty} (also see~\cite{Parikh:1999mf}, for a related treatment).  
Here we use the method to calculate the Hawking 
temperature of a two-dimensional, static and spherically symmetric metric with horizon. We will
then apply this formula to the black hole solution derived above. 

The line element is given by~\eq{ds2d} in the Schwarzschild
gauge $p(r)=1/n(r)$:
\be\label{dsb}
ds^2 = - n(r) dt^2 + n^{-1}(r) dr^2
\,.
\ee
We assume that $n(r)$ vanishes at some $r_0$ and $n'(r)$ is finite and nonzero at $r_0$.
This indicates that there is a horizon at the value $r=r_0$.
One must of course check that the singularity at $r_0$ is only a mathematical one.
In our case this is simple since the only curvature invariant in two-dimensions is $R$, which
remains finite at $r_0$ (see \eq{eqR}).  Expanding $n(r)$ around the
point $r_0$ gives
\be\label{eqr}
n(r) = n'(r_0)(r-r_0) + {\mathcal{O}}[(r-r_0)^2] \equiv {\mathcal{R}}(r_0)(r-r_0)
\,,
\ee
where we assume that ${\mathcal R}(r_0) \neq 0$.  

Now consider a scalar field which satisfies the Klein-Gordon equation
\be
\left(\Box  - \frac{m_0^2}{\hbar^2} \right) \Phi=0
\,.
\ee
If this field is propagating in the background spacetime given by \eq{dsb}, then
\be\label{julia}
-\frac{1}{n(r)}\frac{\d^2 \Phi}{\d t^2} + \frac{\d}{\d r} \left(n(r)\frac{\d \Phi}{\d r} \right) 
= \frac{m_0^2}{\hbar^2} \Phi
\,.
\ee
The semiclassical wavefunctions satisfying the above are
obtained by making the standard ansatz
\be
\Phi(r,t) = e^{\{\frac{i}{\hbar} {\mathcal{S}}(r,t)\}}
\,,
\ee
where $\S$ is a functional which will be expanded in powers of $\hbar$.  Substituting this ansatz into
\eq{julia} gives,
\ba\label{long}
-\frac{1}{n(r)}\left(\frac{\d \S}{\d t}\right)^2 
+ n(r) \left(\frac{\d \S}{\d r}\right)^2 + m_0^2 \nonumber \\
- \frac{i}{\hbar} 
\left[ \frac{1}{n(r)}  \frac{\d^2 \S}{\d t^2} - n(r)  \frac{\d^2 \S}{\d r^2}  
- \frac{dn(r)}{dr} \frac{\d \S}{\d r}      \right] = 0 
\,.
\ea
We now expand $\S$ in a power series of $\hbar/i$:
\be
\S(r,t) = S_0(r,t) + \frac{\hbar}{i} \S_1(r,t) + \left(\frac{\hbar}{i}\right)^2 \S_2(r,t)+\dots
\,,
\ee
and substitute the result into \eq{long}.  Neglecting terms of order $\frac{\hbar}{i}$ or higher we have
\be\label{ss}
-\frac{1}{n(r)}\left(\frac{\d \S_0}{\d t}\right)^2 + n(r) \left(\frac{\d \S_0}{\d r}\right)^2 + m_0^2=0
\,.
\ee
This is simply the Hamilton-Jacobi equation satisfied by a particle of mass $m_0$ moving in a background
spacetime with metric~(\ref{dsb}).  The solution to this equation is
\be
\S_0(r,t) =  -\E t \pm \int^r \frac{dr}{n(r)} \sqrt{\E^2 - m_0^2 n(r)}
\,,
\ee
where $\E$ is a constant and is identified with the energy.  For simplicity we take $m_0=0$.~\footnote{Note
that the results do not change for $m_0 \ne 0$ \cite{Srinivasan:1998ty}.}  Now it
is possible to solve \eq{ss} exactly.  

Using the usual saddle point method, the semiclassical propagator $K(x'',x')$ for a particle propagating
from a spacetime point $x''=(t_1,r_1)$ to a point $x'=(t_2,r_2)$ is
\be
K(x'',x') = \N e^{\frac{i}{\hbar} \S_0(x'',x')}
\,,
\ee
where $\S_0$ is the action functional satisfying the classical Hamilton-Jacobi equation in the massless limit
and $\N$ is a suitable normalization constant.  With the solution to \eq{ss}:
\be\label{sol}
\S_0(x'',x') = -\E(t_2-t_1) \pm \E \int_{r_1}^{r_2} \frac{dr}{n(r)}
\,,
\ee
we can calculate the amplitudes and probabilities of emission and absorption through the event horizon at $r_0$.
Note that the sign ambiguity in \eq{sol} corresponds to the outgoing $(\d \S_0 >0)$ or ingoing $(\d \S_0 <0)$ massless
particles.

If the points $x''$ and $x'$ are on opposite sides of the event horizon then the above integral diverges
(since $n^{-1}$ diverges at $r=r_0$).  Therefore, to evaluate the integral we employ the calculus of residues and choose the
contour over which the integral is to be performed around the point $r=r_0$.

Consider an outgoing particle at $r=r_1 < r_0$.  The modulus squared of the amplitude for this particle
to cross the horizon gives the probability of emission of the particle.  Invoking the usual ``$i \e$" prescription,
the contribution to $\S_0$ in the ranges $(r,r_0 - \e)$ and $(r_0 + \e,r_2)$ is real.  We take the contour to
lie in the upper complex plane and find
\ba
\S_0[emission] = -\E \lim_{\e \rightarrow 0} \int_{r_0 -\e}^{r_0 + \e} \frac{dr}{n(r)} + real\; part \nonumber \\
= \frac{i\pi \E}{{\mathcal R}(r_0)} + real \; part
\,,
\ea
where the minus sign in front of the integral corresponds to the
initial condition that $(\d \S_0 /\d r) >0$ at $r=r_1 < r_0$, and ${\mathcal R}(r_0)$ is given by \eq{eqr}.

Now consider an ingoing particle with $(\d \S_0 /\d r) >0$ at $r=r_2 > r_0$.  The modulus squared of the amplitude
for this particle to cross the horizon gives the probability of absorption of the particle into the horizon.  Choosing the
contour to lie in the upper complex plane gives
\ba
\S_0[absorption] = -\E \lim_{\e \rightarrow 0} \int_{r_0 +\e}^{r_0 - \e} \frac{dr}{n(r)} + real \; part \nonumber \\
= - \frac{i\pi \E}{{\mathcal R}(r_0)} + real \; part
\,.
\ea
This result agrees with the calculation of an outgoing particle $(\d \S_0 /\d r) >0$ at $r=r_2 > r_0$.  Here the contour
is taken in the lower half-plane and the amplitude for the particle to cross the horizon is the same as
that of ingoing particle due to time reversal invariance.

Squaring the modulus to get the probability gives,
\be
P[emission] \propto e^{-\frac{2\pi \E}{\hbar {\mathcal R}}}
\,,
\ee
\be
P[absorption] \propto e^{\frac{2\pi \E}{\hbar {\mathcal R}}}
\,,
\ee
implying that
\be
P[emission] = e^{-\frac{4 \pi \E}{\hbar {\mathcal R}}}P[absorption]
\,.
\ee
Comparing this formula with the relation due to Hawking and Hartle~\cite{Hartle:tp}:
\be
P[emission] = e^{(-\b \E)}P[absorption]
\,,
\ee
we find the identification
\be\label{eqtem}
\beta^{-1} = \frac{\hbar |{\mathcal R}|}{4\pi}
\,.
\ee
The method described above was shown to correctly reproduce the Hawking temperatures of the 
Schwarzschild black hole
\be\label{stemp}
\beta^{-1} = \frac{\hbar}{8\pi m}
\,,
\ee
de Sitter spacetime and Rindler spacetime~\cite{Srinivasan:1998ty}, and is valid for
a variety of coordinate systems~\cite{Shankaranarayanan:2000qv}.  The method employs simple
quantum mechanics and reproduces results obtained by much more difficult means such as
the calculation of Bogoliubov coefficients and other traditional 
techniques, \eg \cite{Hartle:tp},\cite{Hawking:rv},\cite{Hawking:sw}.
\section{An eternally evaporating black hole}
\label{evap}  
Calculating the Hawking temperature of the black hole (given by the metric~(\ref{dsb}),
with~\eq{nofr}) is now a simple task.  Expanding $n(r)$ near $r_0$ gives
\be
n(r)=\frac{2mr_0}{l^2 m + r_0^3} (r-r_0) + {\mathcal{O}}[(r-r_0)^2]
\,.
\ee
In order to agree with the Schwarzschild solution we take $r_0=2 m$,
from which ${\mathcal R}(2 m) = 4m/(l^2 + 8m^2)$.  Hence, the Hawking temperature of 
the black hole is given by (see~\eq{eqtem}),
\be\label{temp}
T_H = \frac{\hbar m}{\pi(l^2 + 8 m^2)}
\,.
\ee
This temperature is plotted in Fig. (3).  
\begin{figure}\label{ht.eps}
\begin{center}
 \epsfxsize=3.2 in \centerline{\epsfbox{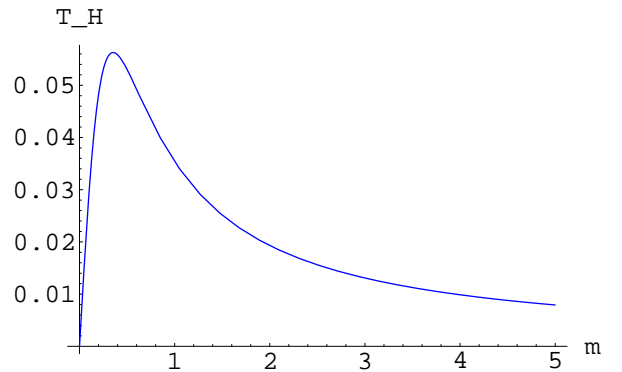}}
  \end{center}
    \caption{The Hawking temperature $T_H$ asymptotically goes to zero as
		the mass goes to zero.}
		    \end{figure}
The most important feature of this solution is that
$T_H \rightarrow 0$ as $m \rightarrow 0$, in contrast to the standard Schwarzschild black hole
which approaches infinite temperature in the zero mass limit (see \eq{stemp}).  Notice that
in the limit of large $m$ (or in the limit as the fundamental parameter $l \rightarrow 0$) the
formula for $T_H$ appropriately reduces to the Schwarzschild temperature.  At
the maximum temperature $T_{max} = \hbar/{4\sqrt{2} \pi l}$, the curvature at the horizon,
remains less than the magnitude of the limiting curvature.  Hence, we are comforted that
our semiclassical treatment of the radiation should remain a good approximation.

\begin{figure}\label{power.eps}
\begin{center}
 \epsfxsize=3.2 in \centerline{\epsfbox{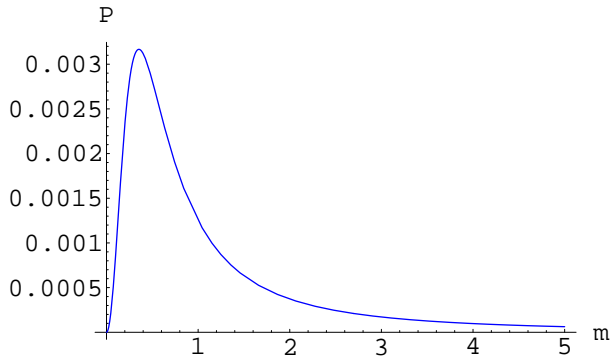}}
  \end{center}
    \caption{The power ${\mathcal{P}} \sim -dm/dt$ versus mass $m$.}
		    \end{figure}

A two dimensional version of Stefan's law gives the total power radiated by the black hole:
\be\label{power}
{\mathcal{P}} \sim - \frac{dm}{dt} \sim T^2
\,.
\ee
The power is plotted in \fig{power.eps}.  
Using the power it is possible to estimate
the evaporation time of the black hole
\be\label{tevap}
t_{evap} \sim \frac{m}{|dm/dt|} \sim \frac{(l^2 + 8m^2)^2}{m}
\,,
\ee
which is infinite in the limit $m\rightarrow 0$.  In the limit of large mass
(or taking $l \rightarrow 0$) the evaporation time reduces to the familiar formula for the
Schwarzschild black hole $t_{evap}\sim m^3$.  Our result implies that the black hole
will radiate eternally.  The mass will decrease as in the Schwarzschild case until most
of the mass is radiated away, at which point the radiation decreases
and one is left with a very slowly radiating, small black hole.~\footnote{Note that a similar behavior
was found in~\cite{Casadio:1997yv}.  Here the authors considered a microcanonical treatment of black holes.}
(At the maximum temperature the black hole
has a Schwarzschild radius of $r_s=l/\sqrt{2}$.)

Note that if this result applies to realistic black holes, then it should significantly
affect the analysis of evaporating black holes created at CERN's Large Hadron Collider (LHC) 
or in future collider experiments (such as CLIC and VLHC).  Most current discussions of
black holes created in the lab are based on semi-classical 
calculations that are valid only when the mass of the black hole is much larger 
than $M_{pl}$.  When the black hole mass approaches the fundamental scale the physics 
required to understand the process of Hawking radiation is rooted in a 
theory of quantum gravity.  Our classical
intuition concerning the creation of black holes in the lab may require refinement. (This is indicated by the
above result).  Further observational consequences of black hole remnants will be discussed in 
a future work~\cite{rhbdae2}.
\section{Miniature black holes as dark matter candidates}
\label{dark}  
One of the greatest challenges of cosmology is to determine the nature of the dark matter/energy 
which makes up most of the matter in the Universe.
The eternal black holes we have described above could have very significant consequences for cosmology, 
since they have the potential to contribute a sizeable amount of the dark matter 
in the Universe today (black hole remnants could behave as weakly interacting matter 
particles (WIMPS)~\cite{Chen:2002tu}).~\footnote{For some early references on black holes as dark
matter candidates see~\cite{Hut:iy,Hegyi:1985tm,Carr:1985tk}.}
It is possible that in the early Universe geometric fluctuations produced a thermal (Boltzmann) distribution of 
black holes (see e.g.,\cite{Kapusta:yh}).  
During the expansion of the Universe these black holes would slowly decay 
into radiation and very small black holes near or below the fundamental scale which would still
be present today.
\section{Conclusions}
\label{conc}  
In this paper we have explored the process of Hawking radiation within 
the context of a specific model for a nonsingular black hole. We find that the black hole
radiation mimics the Schwarzschild case for large values of the mass, but then reaches 
a maximum (determined by the fundamental scale) and then slowly decreases for all values
of the time coordinate. The resulting miniature black holes could play an important role
as dark matter candidates. 

A brief comment on the black hole information loss problem is warranted.
The above model does not suffer from the traditional black hole information loss problem.
All the information which falls into the black hole
is transported into the nonsingular black hole interior. In principle, an observer
travelling into the black hole should be able to recover this information. 
Because the black hole never completely evaporates, the information is never truly lost.

A standard argument against the existence of stable black hole remnants is that it seems
physically unreasonable for a large amount of information to be carried within a small (Planckian)
volume~\cite{Giddings:1992hh}. One possible way to circumvent this issue was proposed within the
context of charged dilatonic black holes~\cite{Callan:rs} - \cite{Giddings:kn}. Here it is
suggested that what appears to be a small volume to an observer outside of the black hole
is actually an infinite volume tube capable of storing infinite information. The scenario
discussed in this paper shares this feature. The black hole is viewed from the outside as
being a small remnant. The large interior core contains the missing information which
may be accessed only by travelling into the black hole. In~\cite{Easson:2001qf} it was argued that the size of the
universe inside the black hole is infinite.
~\footnote{One problem which may remain
in this scenario is that a large number of fundamental-mass particles will appear in loops and
in the thermodynamic ensemble leading to a large degeneracy. It is conceivable, however,
that they could be cured by suppressed amplitudes for creating such large cores if these amplitudes decrease
sufficiently rapidly with the information content~\cite{Giddings:1992hh}.}

Of course, it remains to be shown that such physical properties will carry over to realistic
four-dimensional black holes.
This possibility is currently being investigated~\cite{rhbdae}. In order to generalize the model
to four dimensions we must use a higher derivative theory of gravity.  
The gravitational action should admit a solution which
resembles the Schwarzschild black hole at large distances but with the
singularity replaced by a de-Sitter universe. The method of construction will 
be similar to that in~\cite{Easson:1999xw}.

Our main conclusion is as follows. A specific construction, based on the notion of a limiting curvature,
is capable of removing singularities in cosmological models. In this paper we have argued
that this construction predicts long-lived black hole relics with observational consequences
and possibly solves the information loss problem.
\section*{Acknowledgments}  
I would like to thank R.~Brandenberger, F.~Leblond,  P.~Martineau and R.~Myers for enlightening discussions.  
I would also like to thank the Perimeter Institute in Waterloo, Canada 
for its hospitality during the completion of this work.

\end{document}